\documentclass[conference]{IEEEtran}
\usepackage{amsmath,amssymb,amsfonts,graphicx,cite,hyperref,multirow,color,subfigure}
\def\BibTeX{{\rm B\kern-.05em{\sc i\kern-.025em b}\kern-.08em
    T\kern-.1667em\lower.7ex\hbox{E}\kern-.125emX}}

\begin{document}
\title{Spatial Filtering for Brain Computer Interfaces: \\ A Comparison between the Common Spatial Pattern and Its Variant}

\author{\IEEEauthorblockN{He He and Dongrui Wu, \textit{Senior Member, IEEE}}\\
\IEEEauthorblockA{Key Laboratory of the Ministry of Education for Image Processing and Intelligent Control, \\ School of Automation, Huazhong University of Science and Technology, Wuhan, China}\\
E-mail: hehe91@hust.edu.cn, drwu@hust.edu.cn}

\maketitle

\begin{abstract}
The electroencephalogram (EEG) is the most popular form of input for brain computer interfaces (BCIs). However, it can be easily contaminated by various artifacts and noise, e.g., eye blink, muscle activities, powerline noise, etc. Therefore, the EEG signals are often filtered both spatially and temporally to increase the signal-to-noise ratio before they are fed into a machine learning algorithm for recognition. This paper considers spatial filtering, particularly, the common spatial pattern (CSP) filters for EEG classification. In binary classification, CSP seeks a set of filters to maximize the variance for one class while minimizing it for the other. We first introduce the traditional solution, and then a new solution based on a slightly different objective function. We performed comprehensive experiments on motor imagery to compare the two approaches, and found that generally the traditional CSP solution still gives better results. We also showed that adding regularization to the covariance matrices can improve the final classification performance, no matter which objective function is used.
\end{abstract}


\section{Introduction}

Brain computer interface (BCI) \cite{Wolpaw2002} provides a direct communication pathway between a user and an external device, such that the device can recognize the user's brain status and respond accordingly. BCIs have found successful applications in robotics, text input, games, healthcare, etc \cite{Nicolas-Alonso2012,Erp2012}.

The electroencephalogram (EEG) is the most popular form of BCI input as it is easy to acquire (no surgery) and offers high temporal resolution. However, there are still many challenges for wide-spread real-world applications of EEG-based BCIs \cite{Makeig2012,Lance2012}. An important one is related to the EEG signal quality, as EEG signals can be easily contaminated by various artifacts and noise such as muscle movements, eye blinks, heartbeats, environmental electromagnetic fields, etc. Therefore, it is important to filter the EEG signals both spatially and temporally to increase the signal-to-noise ratio.

This paper focuses on the common spatial pattern (CSP) filters \cite{Koles1990,Muller1999,drwuSF2018,Blankertz2008,Ramoser2000,drwuRG2017,drwuTLCSP2017} for spatial filtering. CSP was first proposed by Koles et al. \cite{Koles1990} to extract discriminative EEG features from two human populations (normal people and patients). Mueller-Gerking et al. \cite{Muller1999} extended it to single-trial classification of motor imagery EEG, which remains the most popular and important application field of CSP filters.

CSP finds a set of spatial filters that can achieve good discrimination among different classes. At the same time, it can reduce the dimensionality of the EEG signals. Different approaches have been proposed in the literature to compute them \cite{Koles1990,Muller1999,drwuSF2018,Blankertz2008}. In this paper we review some typical approaches, and propose a new approach by solving a slightly different objective function in the optimization. We then use two motor imagery datasets in BCI Competition IV to compare these approaches.

The rest of this paper is organized as follows: Section~\ref{sect:csp} introduces existing approaches for computing the CSP filters. Section~\ref{sect:new} introduces the proposed Stiefel manifold based approach. Section~\ref{sect:test} compares the performance of these approaches in motor imagery applications. Finally, Section~\ref{sect:conclusions} draws conclusions.

\section{Common Spatial Pattern (CSP)} \label{sect:csp}

Without loss of generality, this paper considers binary classification only. However, multi-class classification can be extended from binary classification by the one-versus-one approach or the one-versus-the-rest approach \cite{drwuSF2018}.

Let $X\in \mathbb{R}^{C\times T}$ be an EEG epoch, where $C$ is the number of channels and $T$ the number of time samples. Assume Class $c$ has $N_c$ epochs, and $X_{c,i}$ is the $i$th EEG epoch in Class $c$. Then, the class mean covariance matrices are:
\begin{align}
 \Sigma_0&=\frac{1}{N_0}\sum_{i=1}^{N_0}X_{0,i}X_{0,i}^T\\
 \Sigma_1&=\frac{1}{N_1}\sum_{i=1}^{N_1}X_{1,i}X_{1,i}^T
\end{align}

CSP identifies a spatial filtering matrix $W\in\mathbb{R}^{C'\times C}$ ($C'<C$) from $\Sigma_0$ and $\Sigma_1$, which projects the EEG signals from the original $C$-dimensional space to a lower $C'$-dimensional space:
\begin{align}
 X^\ast=W^TX \in\mathbb{R}^{C'\times T}
\end{align}
Each column of $W$ is a spatial filter, and it is designed such that for the projected signal, the variance for one class is maximized, meanwhile the variance for the other class is minimized.

Different motivations and approaches for computing $W$ have been proposed in the literature. Some typical ones are introduced next.

\subsection{Approach 1}

The earliest CSP approach for EEG processing was proposed by Koles et al. \cite{Koles1990}. It first forms a composite covariance matrix:
\begin{align}
 \Sigma=\Sigma_0+\Sigma_1 \label{eq:Sigma}
\end{align}
which has the following singular value decomposition (SVD):
\begin{align}
  \Sigma=U \Lambda U^T \label{eq:SVDsigma}
\end{align}
where $U$ is an orthonormal matrix whose columns are normalized eigenvectors of $\Sigma$, and $\Lambda$ is a diagonal matrix, whose diagonal terms are the corresponding eigenvalues.

It then constructs a whitening matrix
 \begin{align}
 P=\Lambda^{-\frac{1}{2}}U^T \label{eq:P}
\end{align}
to equalize the variances in the space spanned by the eigenvectors. Applying the whitening transformation to both $\Sigma_0$ and $ \Sigma_1$, we have
\begin{align}
  S_0&=P \Sigma_0 P^T \label{eq:S0}\\
  S_1&=P \Sigma_1 P^T \label{eq:S1}\\
  S_0&+S_1=I
\end{align}
where $I$ is the identify matrix.

It can be shown that $S_0$ and $S_1$ share the same eigenvectors, and the sum of the corresponding eigenvalues always equals 1 \cite{Fukunaga1972}, i.e., if we perform eigen decomposition on $S_0$  by
\begin{align}
  S_0&=U \Lambda_0 U^T\label{eq:S0'}
\end{align}
Then we should also have
\begin{align}
  S_1&=U \Lambda_1 U^T \label{eq:S1'}\\
  \Lambda_0&+\Lambda_1=I
\end{align}
where $U$ is an orthogonal matrix whose columns are the normalized eigenvectors, i.e., $UU^T=I$.

Assume the diagonal terms of $\Lambda_0$ has been sorted in descending order. Then, the diagonal terms of $\Lambda_1$ must be in the ascending order. The first column of $U$ accounts for the maximum variance in $S_0$ and the least variance in $S_1$, so it is very useful in discriminating between $S_0$ and $S_1$. Similarly, the last column of $U$ accounts for the least variance in $S_0$ and the maximum variance in $S_1$, so it is also very useful in discriminating between $S_0$ and $S_1$. In practice we usually select a few eigenvectors corresponding to the maximum eigenvalues, and also a few eigenvectors corresponding to the minimum eigenvalues, to form $W$.

In summary, Koles et al. \cite{Koles1990} used the following procedure to compute the CSP matrix $W\in\mathbb{R}^{C'\times C}$:
\begin{enumerate}
\item Compute $\Sigma$ in (\ref{eq:Sigma}) and its SVD in (\ref{eq:SVDsigma}).
\item Compute the whitening matrix $P$ in (\ref{eq:P}).
\item Compute $S_0$ in (\ref{eq:S0}) and its SVD in (\ref{eq:S0'}).
\item Sort the diagonal terms of $\Lambda_0$ in descending order, and adjust the columns of $U$ accordingly. Assemble $V$ as the first and last $C'/2$ columns of $U$.
\item Compute $W=P^TV$.
\end{enumerate}

\subsection{Approach 2}

In Approach 1, we need to first calculate the whitening matrix $P$, then the orthogonal transformation matrix $U$, assemble $V$ from $U$, and finally obtain $W=P^TV$. Blankertz et al. \cite{Blankertz2008} proposed a simpler solution to compute $W$ directly from $\Sigma_1$ and $\Sigma_2$.

According to \cite{Fukunaga1972}, we can simultaneously diagonalize the two mean covariance matrices:
\begin{align}
  V^T \Sigma_0 V&=\Lambda_0\\
  V^T \Sigma_1 V&=\Lambda_1
\end{align}
where $\Lambda_0+\Lambda_1=I$.

Assume the diagonal terms of $\Lambda_0$ has been sorted in descending order. Then, the diagonal terms of $\Lambda_1$ must be in the ascending order. The first a few columns of $V$ account for the maximum variance in $\Sigma_0$ and the least variance in $\Sigma_1$, so they are very useful in discriminating between $\Sigma_0$ and $\Sigma_1$. Similarly, the last a few columns of $V$ account for the least variance in $\Sigma_0$ and the maximum variance in $\Sigma_1$, so they are also very useful in discriminating between $\Sigma_0$ and $\Sigma_1$.

In summary, Blankertz et al. \cite{Blankertz2008} used the following procedure to compute CSP matrix $W\in\mathbb{R}^{C'\times C}$:
\begin{enumerate}
\item Solve the following generalized eigenvalue problem
\begin{align}
  \Sigma_0 \mathbf{w}=\lambda \Sigma_1 \mathbf{w} \label{eq:GE}
\end{align}
to obtain $\lambda_i$ and the corresponding $\mathbf{w}_i$, $i=1,...,C$.
\item Sort $\lambda_i$ and the corresponding $\mathbf{w}_i$ in descending order.
\item Construct $W=[\mathbf{w}_1,...,\mathbf{w}_{\frac{C'}{2}},\mathbf{w}_{C-\frac{C'}{2}+1},\mathbf{w}_C]$, i.e., $W$ uses the first and last $\frac{C'}{2}$ $\mathbf{w}_i$ as its columns.
\end{enumerate}

\subsection{Discussions}

Although Approaches 1 and 2 use different procedures to compute $W$, they actually give the same results.

For Approach 1, from (\ref{eq:S0}) and (\ref{eq:S0'}) we have:
\begin{align}
 S_0= P \Sigma_0 P^T=U \Lambda_0 U^T
\end{align}
which can be rewritten as:
\begin{align}
 \Sigma_0 P^TU=  P^{-1}U \Lambda_0 \label{eq:Dis2}
\end{align}

Since $P$ is the whitening matrix for the composite covariance matrix $\Sigma$, i.e.,
\begin{align}
  P \Sigma P^T=I
\end{align}
we have
\begin{align}
  P^{-1}= \Sigma P^T \label{eq:P-1}
\end{align}

Substituting (\ref{eq:P-1}) into (\ref{eq:Dis2}), it follows that:
\begin{align}
 \Sigma_0 P^TU=\Sigma P^TU \Lambda_1
\end{align}
i.e.,
\begin{align}
\Sigma^{-1}\Sigma_0(P^TU)= (P^TU) \Lambda_1
\end{align}
So the filtering matrix $P^TU$ consists of the eigenvectors of $\Sigma^{-1}\Sigma_0$, which are identical to the eigenvectors of $\Sigma_1^{-1}\Sigma_0$. $W=P^TV$ consists a subset of $P^TU$.

For Approach~2, (\ref{eq:GE}) can be rewritten as:
\begin{align}
\Sigma_1^{-1}\Sigma_0\mathbf{w}=\lambda \mathbf{w}
\end{align}
So each $\mathbf{w}$ is also an eigenvector of $\Sigma_1^{-1}\Sigma_0$.

In summary, the above derivations show that $W$ can be constructed from the eigenvectors of $\Sigma_1^{-1}\Sigma_0$, and this is actually the most frequently used approach in computing the CSP in practice.

Finally, if we take a closer look at the rationale for CSP, we can conclude that the CSP filters actually optimize the following objective function:
\begin{align}
Ratio1=\arg\max_{W=[\mathbf{w}_1,...,\mathbf{w}_{C'}]}
&\sum_{i=1}^{C'/2} \frac{\mathbf{w}_i^T\Sigma_0\mathbf{w}_i}{\mathbf{w}_i^T\Sigma_1\mathbf{w}_i}\nonumber \\
&+\sum_{i=C'/2+1}^{C'} \frac{\mathbf{w}_i^T\Sigma_1\mathbf{w}_i}{\mathbf{w}_i^T\Sigma_0\mathbf{w}_i} \label{eq:R1}
\end{align}

\section{A New Approach for Computing the Spatial Filters in the Stiefel Manifold} \label{sect:new}

$Ratio1$ in (\ref{eq:R1}) optimizes the sum of ratios. A closely related objective function is to optimize the ratio of sums:
\begin{align}
Ratio2=\arg\max_{W=[\mathbf{w}_1,...,\mathbf{w}_{C'}]}
&\frac{\sum_{i=1}^{C'/2} \mathbf{w}_i^T\Sigma_0\mathbf{w}_i}{\sum_{i=1}^{C'/2} \mathbf{w}_i^T\Sigma_1\mathbf{w}_i}\nonumber \\
&+\frac{\sum_{i=C'/2+1}^{C'} \mathbf{w}_i^T\Sigma_1\mathbf{w}_i}{\sum_{i=C'/2+1}^{C'} \mathbf{w}_i^T\Sigma_0\mathbf{w}_i} \label{eq:R2}
\end{align}
We are interested in (\ref{eq:R2}) because:
\begin{enumerate}
\item (\ref{eq:R2}) is very similar to (\ref{eq:R1}). In fact, they are identical when $C'=2$. So, we would like to investigate whether this new objective function could result in better classification performance.
\item Although both $\Sigma_0$ and $\Sigma_1$ are symmetric, generally $\Sigma_1^{-1}\Sigma_0$ is not symmetric, so its eigenvectors are not mutually orthogonal. In other words, the columns of $W$ obtained from optimizing $Ratio1$ are correlated, which may encode redundant information. On the other hand, when solving $Ratio2$, it is possible to make the first $C'/2$ columns of $W$ orthogonal, and also the last $C'/2$ columns orthogonal (although the first and last $C'/2$ columns are generally not mutually orthogonal). It's interesting to investigate whether this orthogonality can improve the classification performance.
\end{enumerate}

The two terms in (\ref{eq:R2}) are independent, so we can compute them separately. Unfortunately, they do not have a closed-form solution. Cunningham and Ghahramani \cite{Cunningham2015} proposed an approach for solving this problem in the Stiefel manifold (SM), which is the set of ordered tuples of orthonormal vectors. Their algorithm is complex and iterative, and we refer the readers to \cite{Cunningham2015} for it. The authors have also provided their Matlab code\footnote{\href{http://github.com/cunni/ldr}{http://github.com/cunni/ldr}}, which was used in our experiment.

\section{Experiments} \label{sect:test}

In this section we compare the SM approach with the traditional CSP approach on two different motor imagery datasets, using two classifiers.

\subsection{Datasets}

Both datasets were from BCI Competition IV\footnote{\href{http://www.bbci.de/competition/iv/}{http://www.bbci.de/competition/iv/}.}.

The first is Dataset 1 \cite{Blankertz2007}, which was recorded from seven healthy subjects. For each subject two classes of motor imagery were selected from the three classes: left hand, right hand, and foot. Continuous EEG signals were acquired from 59 channels and were divided into three parts: calibration data, evaluation data, and special feature. We only used calibration data in this paper, and each subject had 100 trials in each class.

The second is Dataset 2a, which consists of EEG data from nine subjects. Every subject was instructed to perform four different motor imagery tasks, namely the imagination of movement of the left hand, right hand, both feet, and tongue. The signals were recorded using 22 EEG channels and 3 EOG channels. We only used two classes (left hand and right hand), and each class has 72 trials.

\subsection{Preprocessing and Classifiers}

The EEG signals were preprocessed using the Matlab EEGLAB toolbox \cite{Delorme2004}, following the guideline in \cite{Blankertz2008}. First, a band-pass filter (7-30 Hz) was applied to remove muscle artifacts, line-noise contamination and DC drift. Then, we extracted EEG signals between [1, 3.5] seconds after the cue appearance as our trials.

For each subject, we randomly selected 50\% trials for training, and the remaining 50\% for testing, and repeated this process 30 times to get statistically meaningful results. For a given partition, we computed the spatial filters by the traditional CSP approach, and also the proposed SM approach. We then tested two different classifiers:
\begin{enumerate}
\item Linear discriminant analysis (LDA), as in \cite{Blankertz2008}. The features for the $i$th trial were:
\begin{align}
    f_i^j=\log(\mathbf{w}_j^TX_iX_i^T\mathbf{w}_j), \quad j=1,...,C'
\end{align}

\item Minimum distance to Riemannian mean (MDRM), as in \cite{Barachant2012}. The features were the covariance matrices of the trials.
\end{enumerate}

\subsection{Experimental Results}

\begin{figure}[htpb]\centering
\subfigure[]{\includegraphics[width=.95\linewidth,clip]{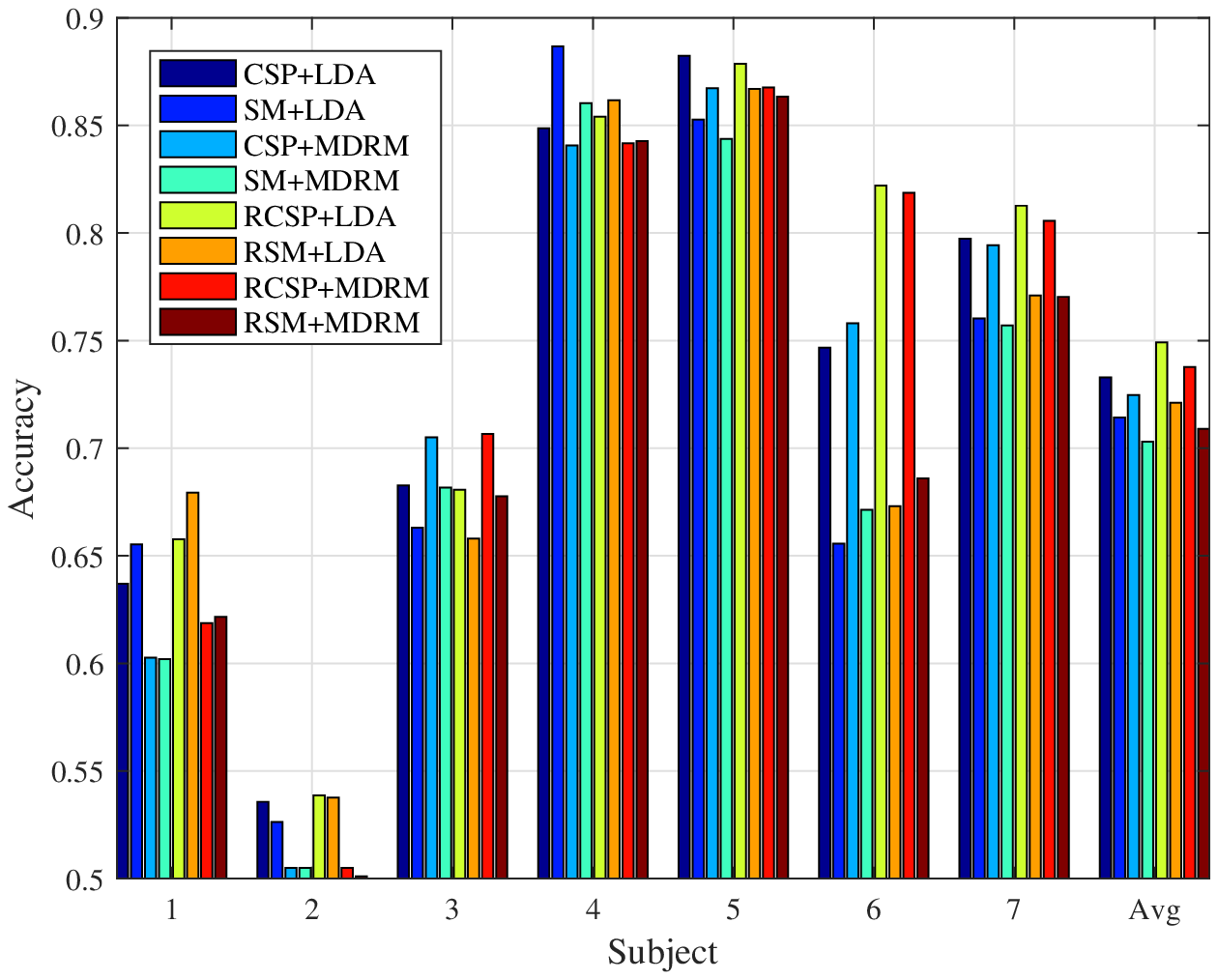}} \label{fig:Acc4}
\subfigure[]{\includegraphics[width=.95\linewidth,clip]{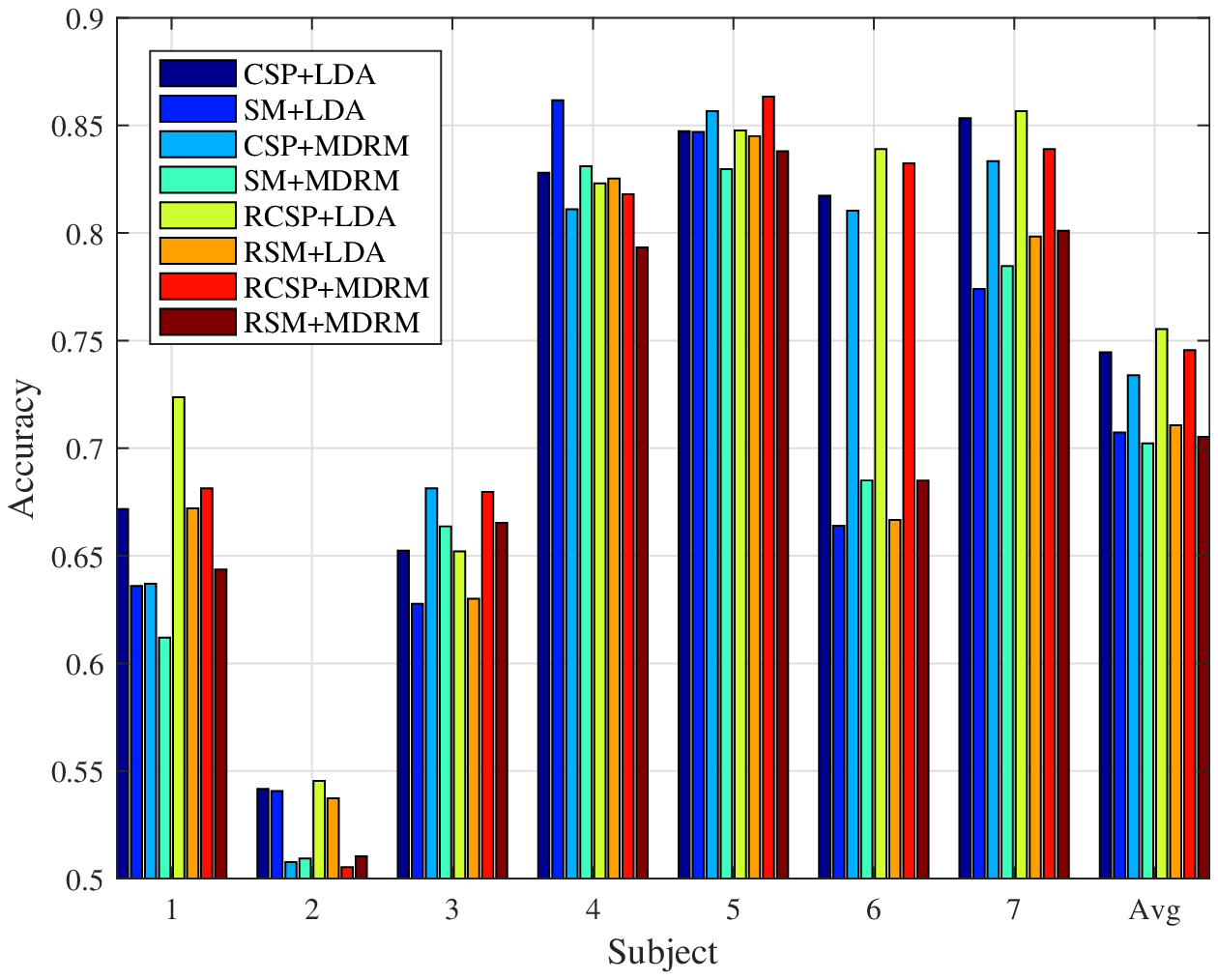}} \label{fig:Acc6}
\subfigure[]{\includegraphics[width=.95\linewidth,clip]{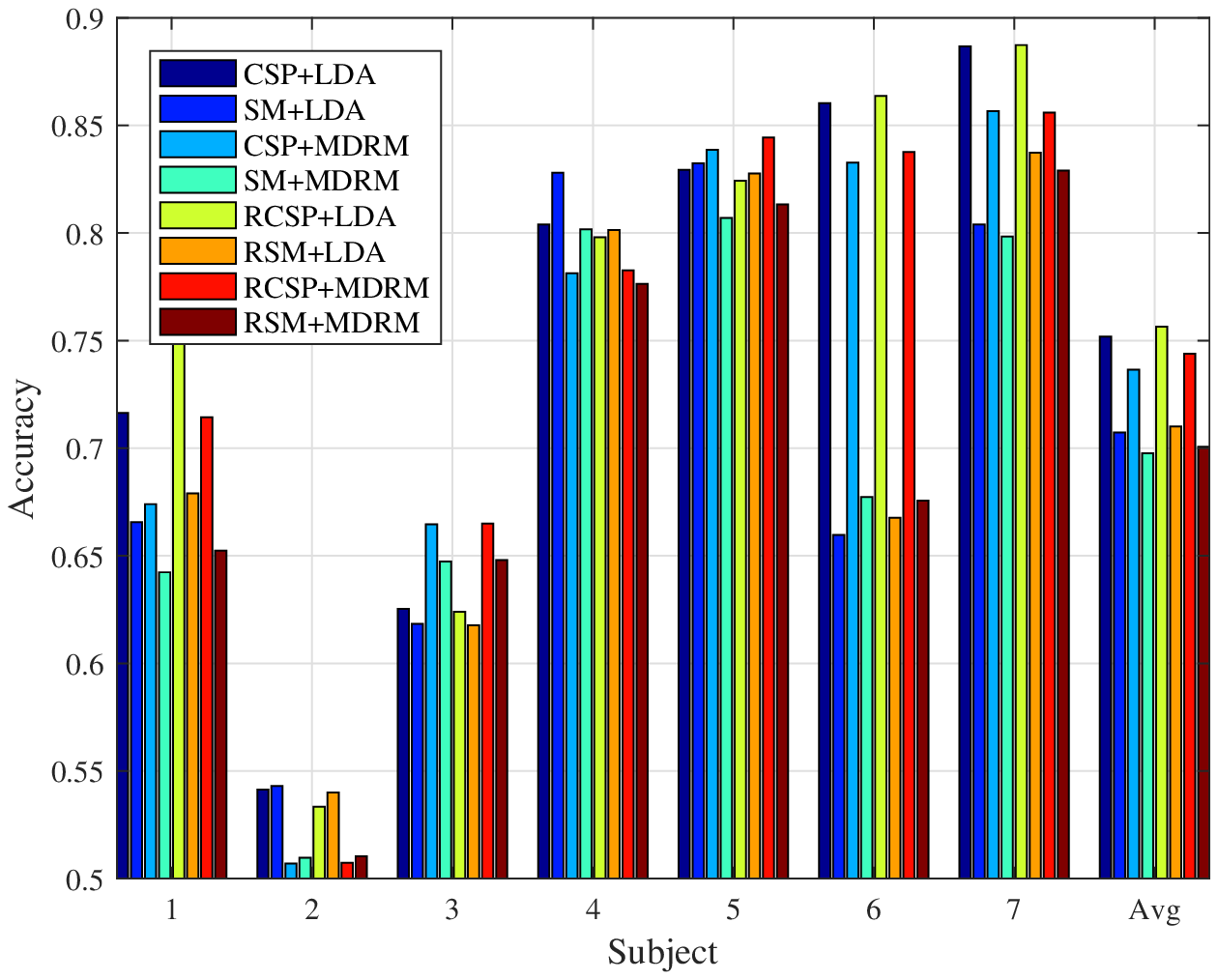}} \label{fig:Acc8}
\caption{Classification accuracies on Dataset 1. (a) $C'=4$; (b) $C'=6$; (c) $C'=8$.}\label{fig:DS1Acc}
\end{figure}

\begin{figure}[htpb]\centering
\subfigure[]{\includegraphics[width=.95\linewidth,clip]{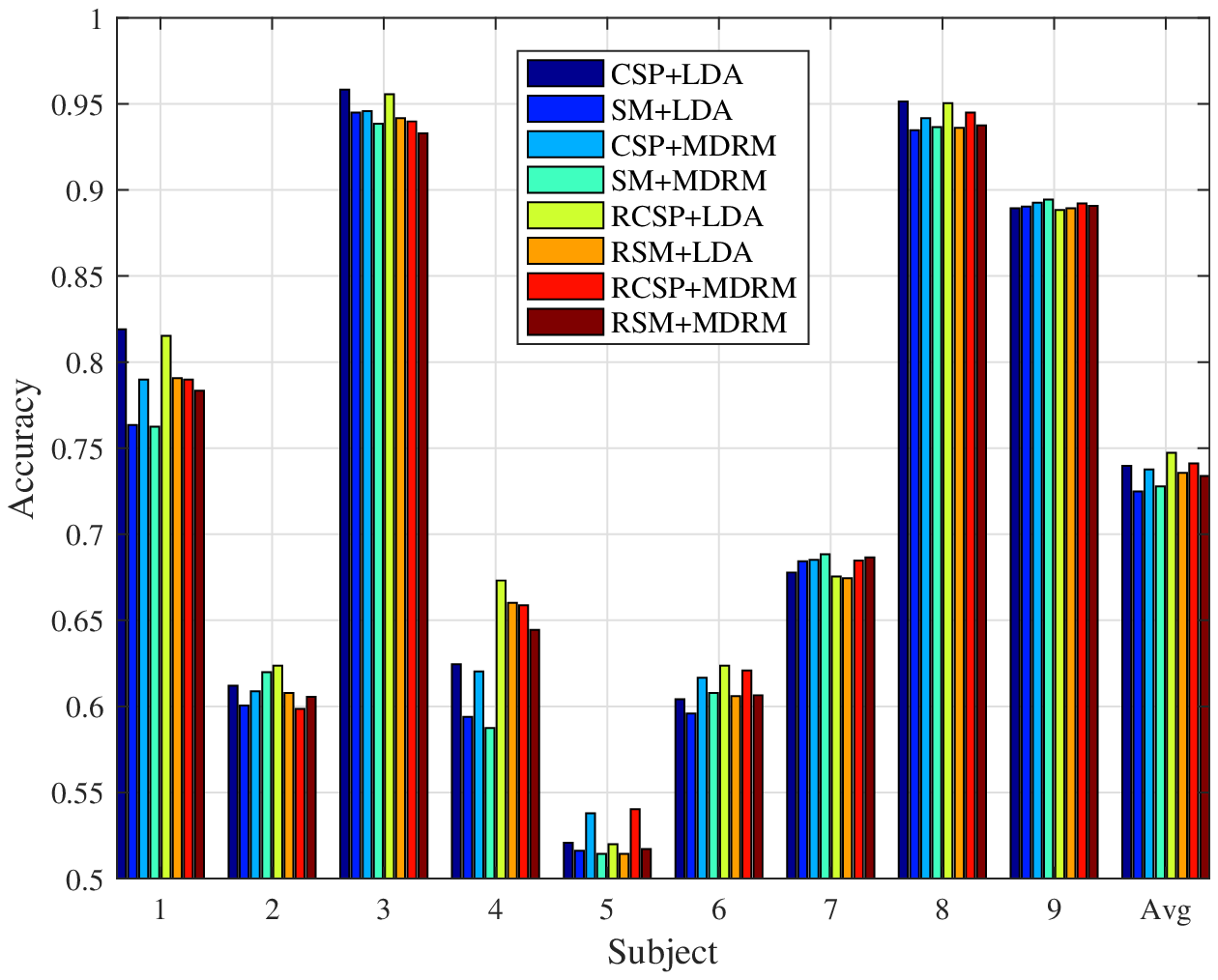}} \label{fig:DS2Acc4}
\subfigure[]{\includegraphics[width=.95\linewidth,clip]{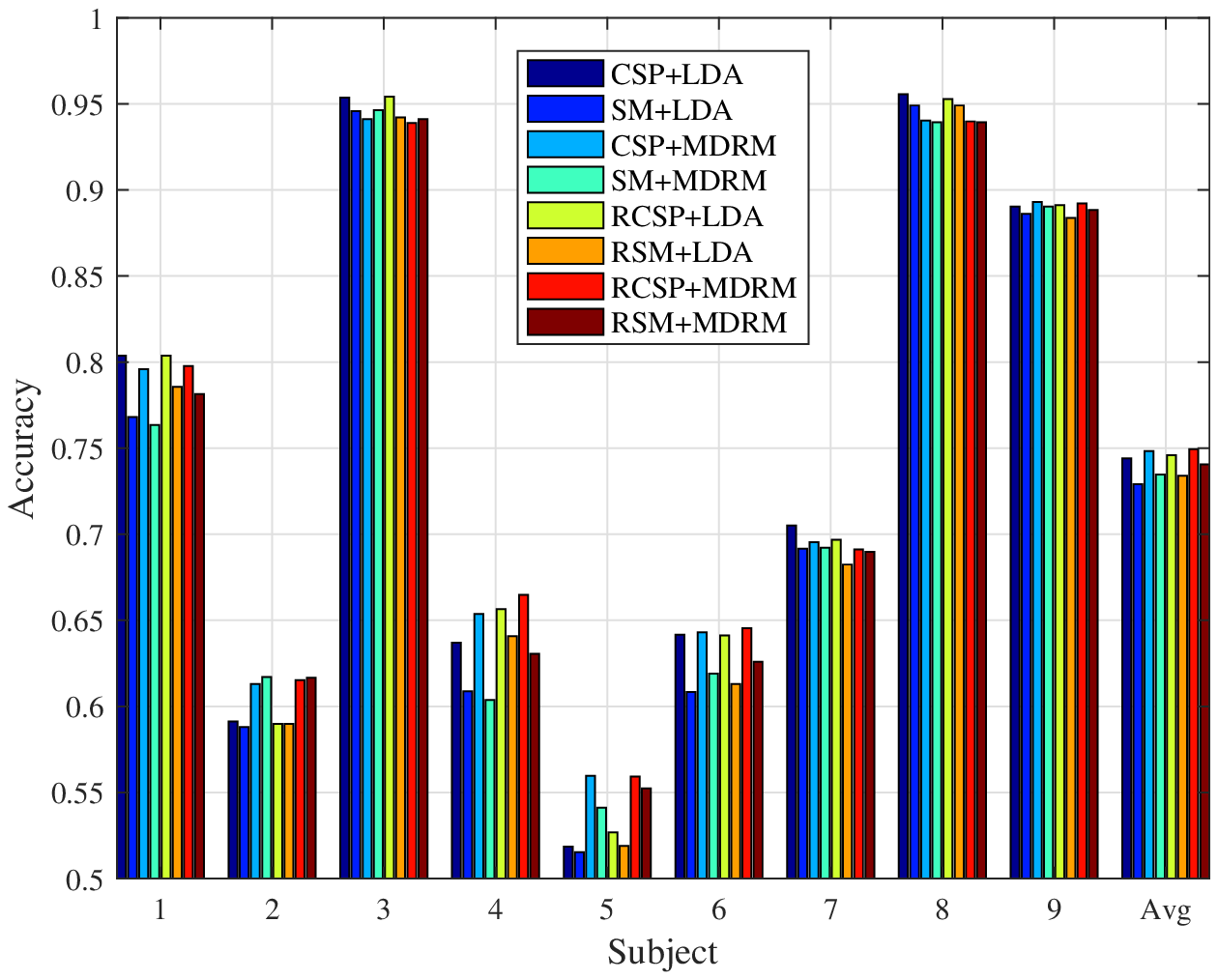}} \label{fig:DS2Acc6}
\subfigure[]{\includegraphics[width=.95\linewidth,clip]{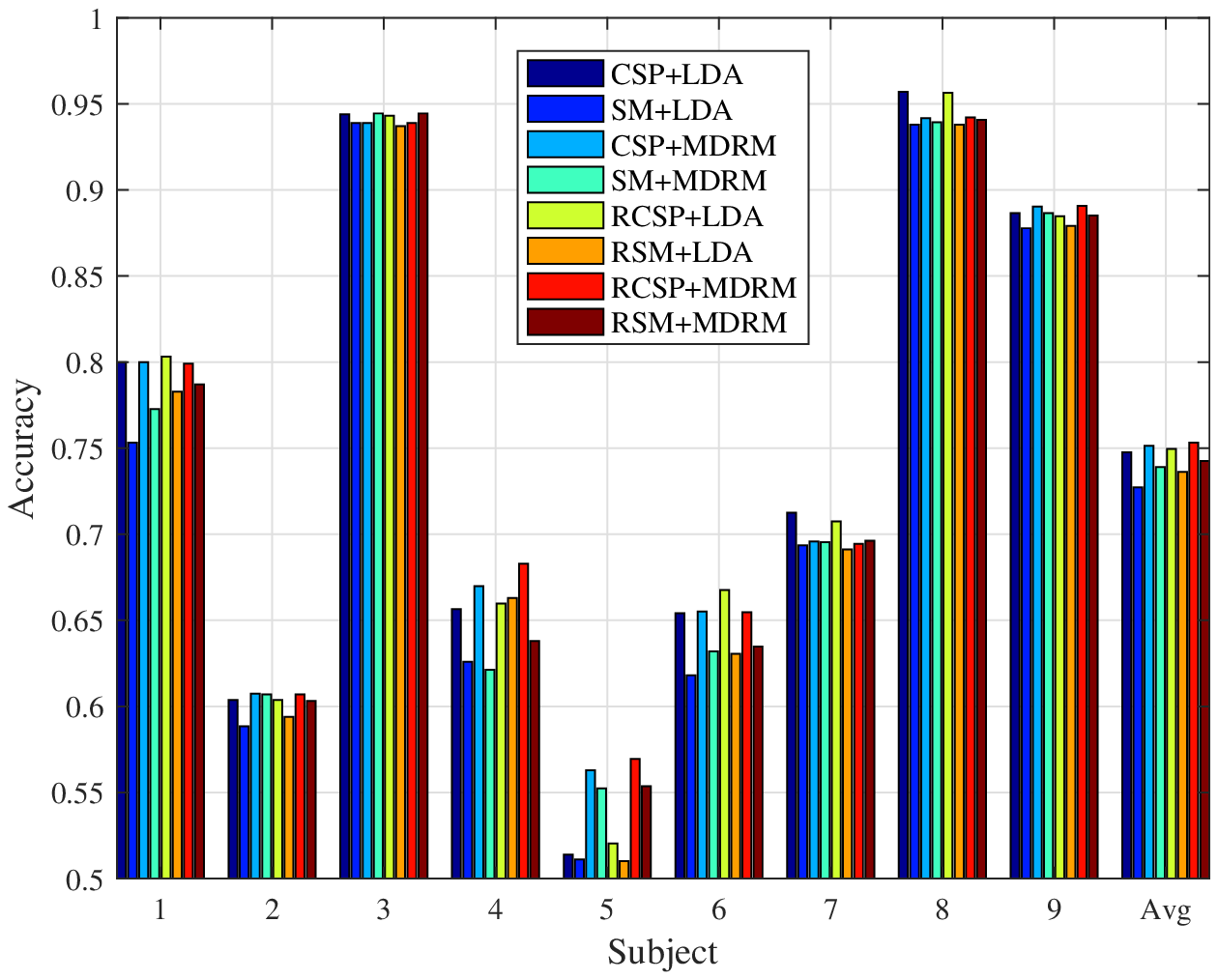}} \label{fig:DS2Acc8}
\caption{Classification accuracies on Dataset 2. (a) $C'=4$; (b) $C'=6$; (c) $C'=8$.}\label{fig:DS2Acc}
\end{figure}

The classification accuracies for different subjects, averaged across 30 runs, are shown as the first four bars in Figs.~\ref{fig:DS1Acc} and \ref{fig:DS2Acc}, for different number of spatial filters. The horizontal axis shows the indices of the subjects, and also the average across the subjects. Observe that for both LDA and MDRM, generally the performances of SM were slightly worse than CSP. Also, generally the performance of MDRM was slightly worse than LDA.

\begin{figure}[htpb]\centering
\subfigure[]{\includegraphics[width=.9\linewidth,clip]{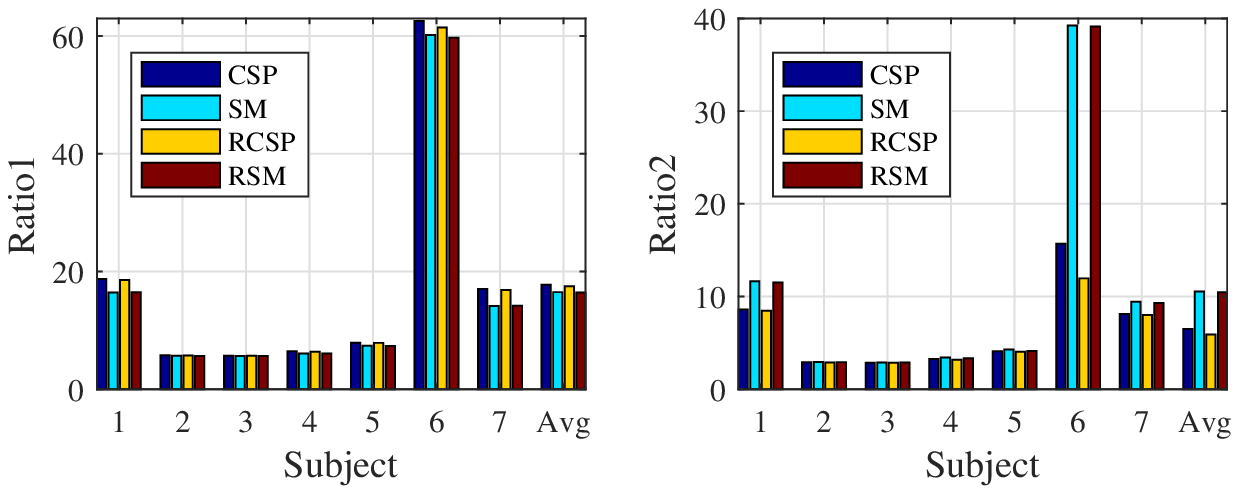}} \label{fig:R4}
\subfigure[]{\includegraphics[width=.9\linewidth,clip]{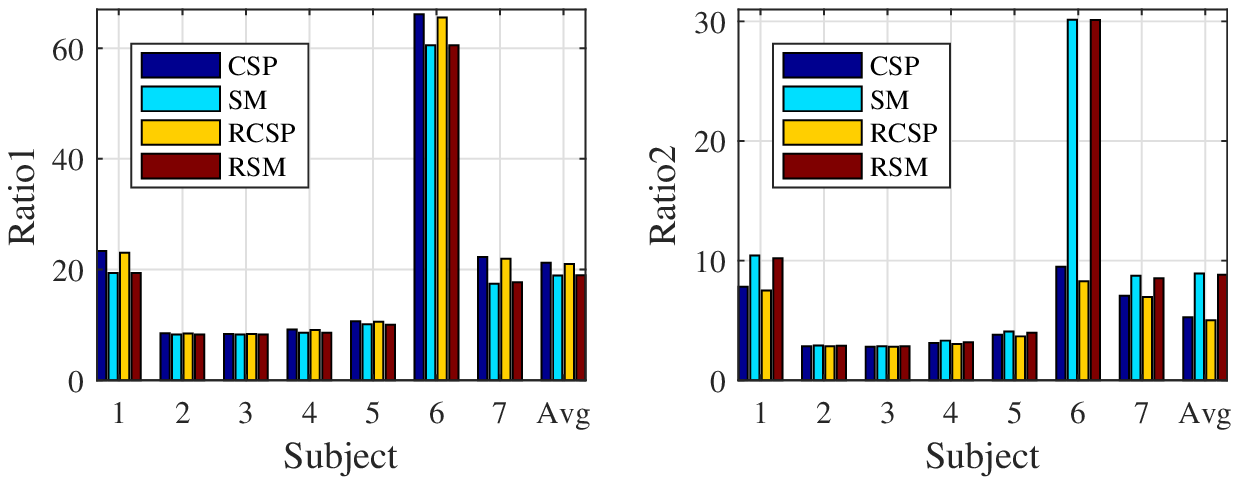}} \label{fig:R6}
\subfigure[]{\includegraphics[width=.9\linewidth,clip]{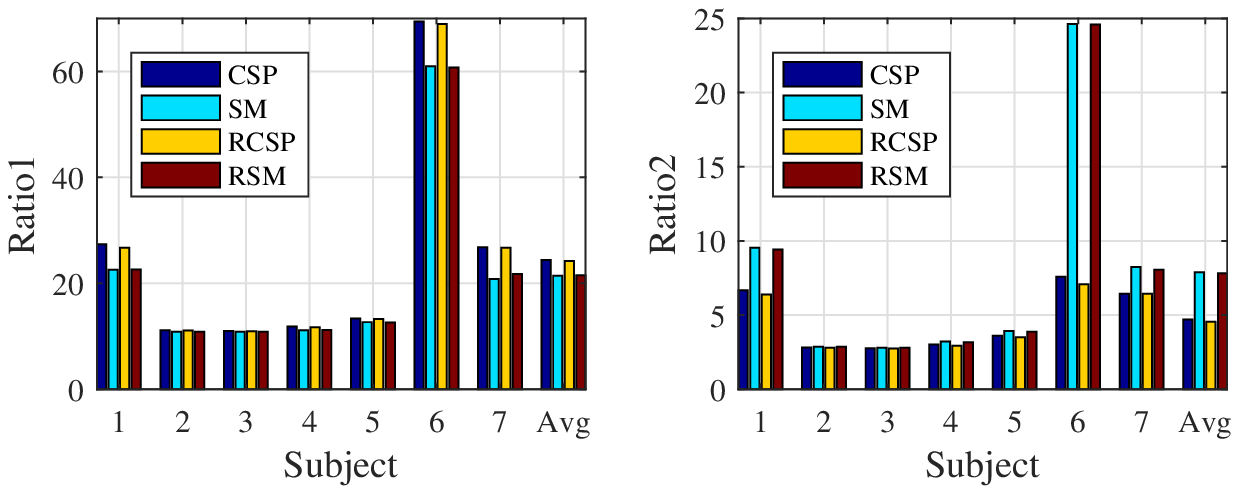}} \label{fig:R8}
\caption{$Ratio1$ and $Ratio2$ on Dataset 1. (a) $C'=4$; (b) $C'=6$; (c) $C'=8$.}\label{fig:DS1R}
\end{figure}

\begin{figure}[htpb]\centering
\subfigure[]{\includegraphics[width=.9\linewidth,clip]{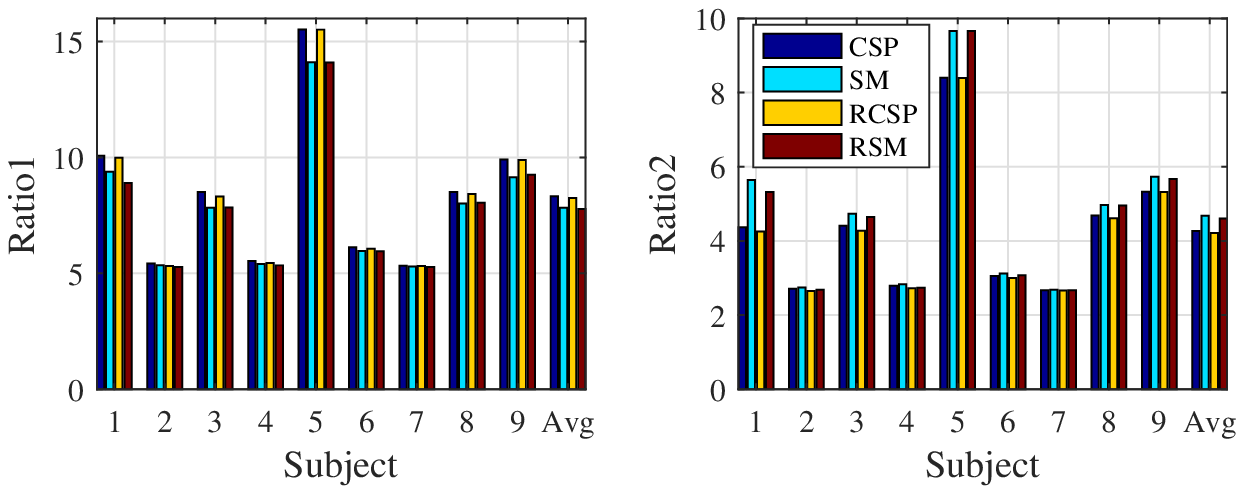}} \label{fig:DS2R4}
\subfigure[]{\includegraphics[width=.9\linewidth,clip]{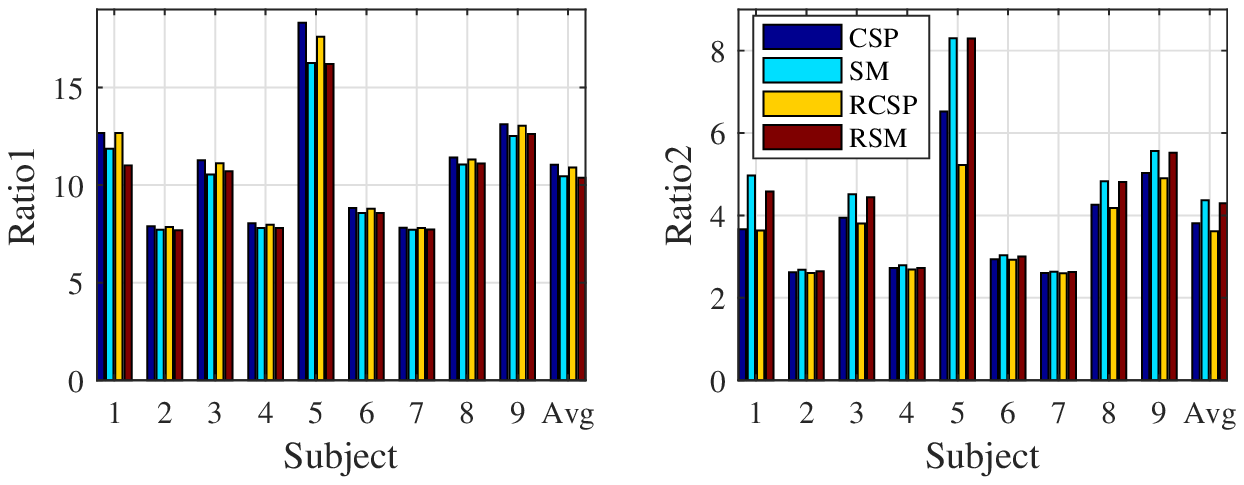}} \label{fig:DS2R6}
\subfigure[]{\includegraphics[width=.9\linewidth,clip]{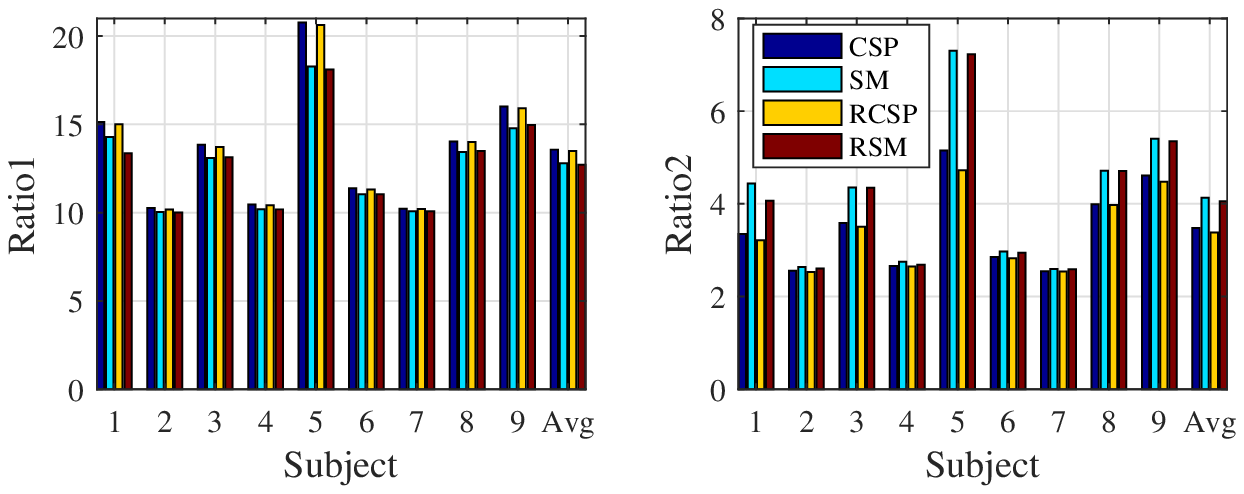}} \label{fig:DS2R8}
\caption{$Ratio1$ and $Ratio2$ on Dataset 2. (a) $C'=4$; (b) $C'=6$; (c) $C'=8$.}\label{fig:DS2R}
\end{figure}

$Ratio1$ and $Ratio2$ from the two objective functions are shown in Figs.~\ref{fig:DS1R} and \ref{fig:DS2R}. Observe that CSP always had higher $Ratio1$ than SM, and SM always had higher $Ratio2$ than CSP, which are as expected. However, it seems that $Ratio1$ is a better objective function, since a higher $Ratio1$ usually results in a better classification accuracy.

The average correlation coefficients between the columns of $W$ for different subjects are shown in Tables~\ref{tab:CC1} and \ref{tab:CC2}, for Datasets 1 and 2, respectively. Observe that on average the columns of $W$ computed from the SM approach were less correlated; however, this did not necessarily result in better classification performance.

\begin{table}[htpb] \centering \setlength{\tabcolsep}{1mm}
\caption{Average correlation coefficients between the columns of $W$ for Dataset 1.}   \label{tab:CC1}
\begin{tabular}{c|cccccccc}   \hline
Subject & 1 & 2 & 3 & 4 & 5 & 6 & 7 & Avg \\ \hline
CSP &    .1393 &   .0760  &  .1064  &  .0797 &   .0839  &  .1122  &  .1148 & .1018\\
SM &     .0907  &  .0477   & .0399   & .1373  &  .0738   & .0943   & .0886 & .0817\\   \hline
\end{tabular}
\end{table}

\begin{table}[htpb] \centering \setlength{\tabcolsep}{.5mm}
\caption{Average correlation coefficients between the columns of $W$ for Dataset 2.}   \label{tab:CC2}
\begin{tabular}{c|cccccccccc}   \hline
Subject & 1 & 2 & 3 & 4 & 5 & 6 & 7 & 8 & 9 & Avg\\ \hline
CSP &     .1807  &   .3224  &   .2571  &   .1776   &  .1416   &  .2116  &   .2419  &   .1904   &  .1665 & .2100\\
SM &      .1744 &    .1759 &    .1522 &    .1488  &   .1280  &   .1481 &    .1354 &    .1288  &   .1521 & .1493 \\   \hline
\end{tabular}
\end{table}

\subsection{Discussions}

The above results showed that generally a larger $Ratio1$ results in better classification performance. We would like to study if this is always true. Lotte and Guan \cite{Lotte2011a} showed that regularization on the traditional CSPs can improve the classification performance. In this subsection we study if regularization can also improve $Ratio1$, i.e., the improved performance is due to the increased $Ratio1$.

Several different regularized CSP (RCSP) approaches have been proposed in \cite{Lotte2011a}. In this paper we compute the first $C'/2$ columns of $W$ in the RCSP from the eigenvectors of $(\Sigma_1+\lambda I)^{-1}\Sigma_0$, and the last $C'/2$ columns from the eigenvectors of $(\Sigma_0+\lambda I)^{-1}\Sigma_1$, where $\lambda$ is an adjustable parameter identified by cross-validation on the labeled data \cite{Lotte2011a}. This approach has showed good performance in \cite{Lotte2011a}.

Similarly, we also develop a regularized SM (RSM) approach, where the first $C'/2$ columns of $W$ in the RSM are computed from maximizing $\frac{\sum_{i=1}^{C'/2} \mathbf{w}_i^T\Sigma_0\mathbf{w}_i}{\sum_{i=1}^{C'/2} \mathbf{w}_i^T(\Sigma_1+\lambda I)\mathbf{w}_i}$, and the last $C'/2$ columns are computed from maximizing $\frac{\sum_{i=C'/2+1}^{C'} \mathbf{w}_i^T\Sigma_1\mathbf{w}_i}{\sum_{i=C'/2+1}^{C'} \mathbf{w}_i^T(\Sigma_0+\lambda I)\mathbf{w}_i}$, in which $\lambda$ is again an adjustable parameter identified by cross-validation on the labeled data.

The classification accuracies for different subjects, averaged across 30 runs, are shown as the last four bars in Figs.~\ref{fig:DS1Acc} and \ref{fig:DS2Acc}, for different number of spatial filters. Observe that:
\begin{enumerate}
\item Generally the performance of RCSP was better than CSP, and the performance of RSM was better than SM, for both LDA and MDRM. This confirmed the observations in \cite{Lotte2011a}.
\item For both LDA and MDRM, generally the performances of RSM were slightly worse than RCSP. This pattern was also observed between SM and CSP.
\item For both RCSP and RSM, generally the performance of MDRM was slightly worse than LDA. Again, this pattern was observed before for CSP and SM.
\end{enumerate}

Next we check if the improved performance of RCSP and RSM over CSP and SM was indeed due to increased $Ratio1$. For this purpose, we plot $Ratio1$ and $Ratio2$ from RCSP and RSM as the last four bars in Figs.~\ref{fig:DS1R} and \ref{fig:DS2R}. Observe that:
\begin{enumerate}
\item Although RCSP had higher classification accuracy than CSP, its $Ratio1$ and $Ratio2$ were slightly worse than those from the unregularized CSP. A similar pattern can also be observed from RSM and SM. This is reasonable, as CSP directly optimizes $Ratio1$, whereas the objective function for RCSP is slightly different.
\item RCSP always had higher $Ratio1$ than RSM, and RSM always had higher $Ratio2$ than RCSP. This pattern was similar to what we have observed on CSP and SM.
\end{enumerate}

From all these observations, we can reach the following two conclusions:
\begin{enumerate}
\item Although the SM approach optimizes an objective function very similar to the objective function of the traditional CSP, it generally results in slightly worse classification performance. Meanwhile, the SM approach does not have a closed-form solution, and hence requires much higher computational cost. For these two reasons, the new objective function and optimization method are not recommended in designing spatial filters.
\item $Ratio1$ in general is a reliable performance measure for the spatial filters. However, because $\Sigma_0$ and $\Sigma_1$ may be noisy, adding regularization to the covariance matrices can improve the final classification performance, although it may reduce $Ratio1$ very slightly. This suggests that maybe it is possible to define an improved objective function for the spatial filters, which is better correlated with the final classification performance. This is one of our future research directions.
\end{enumerate}

\section{Conclusions} \label{sect:conclusions}

CSP is a popular spatial filtering approach for EEG-based BCIs, especially for motor imagery applications. It is used to increase the signal-to-noise ratio of EEG signals before they are fed into a classifier. Its main idea is to project the EEG signals from the original sensor space into a low-dimensional space which maximizing the variance for one class while minimizing it for the other. Different motivations and approaches have been proposed to compute the CSP filters. In this paper, we gave an overview of some typical approaches, and showed that they lead to the same closed-form solution. We also proposed a new objective function, which closely resembles the objective function of the traditional CSP, for developing spatial filters. Experimental results on two Motor Imagery datasets showed that the new objective function results in slightly worse spatial filters than the CSP filters, so the traditional CSP approach is still preferred in practice. Moreover, we also confirmed that adding regularization on the covariance matrices can improve the classification performance, no matter which objective function is used.


\end{document}